\newcommand{\bee}{\begin{equation}}
\newcommand{\ene}{\end{equation}}
\newcommand{\beea}{\begin{eqnarray}}
\newcommand{\enea}{\end{eqnarray}}

\documentclass[aps,preprint,pre,showpacs,superscriptaddress]{revtex4}
\usepackage{amsmath} 
\usepackage{graphicx}
\usepackage{dcolumn}
\usepackage{bm}
\begin{document}
\title{Pseudo-Resonance and Energy Band Gap in Plasmonic Crystals}
\author{M. Akbari-Moghanjoughi}
\affiliation{Faculty of Sciences, Department of Physics, Azarbaijan Shahid Madani University, 51745-406 Tabriz, Iran}

\begin{abstract}
In this paper, using the generalized coupled pseudoforce model with driving elements, we develop a method to study the plasmon excitations and energy band structure in a plasmonic crystal. It is shown that the presence of the periodic ion core potential leads to pseudo-resonance condition in the plasmon wavefunction and electrostatic potential profiles, quite analogous to the frequency resonance, leading to the gap formation in the energy dispersion profiles. It is found that the dual length scale character of plasmon excitations lead to occurrence of a critical value of $a_c=2\pi\lambda_p$ for the lattice constant ($\lambda_p$ being the plasmon wavelength) above and below which the energy band structure of plasmonic crystals becomes substantially different. It is also found that energy band gap positions of parabolic free electron energy dispersion relation are slightly higher compared to that of the plasmon excitations. Here based on the plasmon definition and due to the dual length scale character of plasmons compared to that of single electrons, we provide a simplified interpretation of the wave-particle duality in quantum physics and provide the length scale regime in which both wave and particle properties can be simultaneously examined in a single experiment, where, the complementarity principle breaks down. \textbf{However, the dual characteristics of electron-plasmon coupled excitations, quite analogous to starling murmuration phenomenon, strongly suggests that the long south de Broglie-Bohm pilot wave of electron is nothing but the plasmon itself.} The wavefunction and electrostatic potential solution for a one dimensional plasmonic lattice with a generalized periodic potential is also derived in this research. Current development helps to illuminate the basic properties of quantum phenomenon in many physical context by appropriately incorporating the collective effect in the root level and can have a wide range of applications in the rapidly growing fields of nano-electronics and plasmonics.
\end{abstract}

\date{\today}

\maketitle

\section{Historical Background}

The nineteenth century started with a full range invention of the modern physics following the pioneering photon solution of Max Planck to the half a century old black body radiation problem \cite{planck}. The wavy look at the subatomic world was the next milestone in the new era of modern physics which started by the invention of the most celebrated equation of the physics history by Erwin Schr\"{o}dinger \cite{es} in 1926. Although the new idea of quantum mechanics and its interpretation based on Heisenberg uncertainty principle \cite{hb} was too wavy to be accepted by Albert Einstein, the sceptic father of modern large scale physics theory of gravitation \cite{en}, subsequent extensive rapid theoretical developments \cite{born1,born2,bohr} followed by many ground breaking experiments \cite{dg} put the theory in a firmer ground of almost all in the history of theoretical foundation of physics. Despite the overwhelming success of the new theory in description of microscopic phenomena, however, the somehow disturbing blurry wave-particle nature of physical entities led some founders of the theory like Louis de Broglie \cite{de} to propose the deterministic pilot wave version of the quantum theory which was at the sharp contrast with the early Copenhagen statistical interpretation of quantum mechanics by Max Born \cite{born3}. While the new theory of pilot wave was ignored at the time, it was later reestablished by Bohm \cite{bohm1} in 1952. Many recent pioneering experiments with millimeter-sized walking droplets on a suitably driven fluid surface \cite{co1,co2,bush,co3} have confirmed the existence of variety of macroscopic level quantum features. These droplets have been rigorously shown to mimic variety of quantum-like effects which were already supposed to emerge at the atomic or molecular scales. Prominent classical quantum features range from the most basic interference effect for droplets passing through single and double slit barriers \cite{co4} quite reminiscent of quantum entities, up to the quantum tunneling effect \cite{eddi}, bound state orbital quantization \cite{fort1} and the Landau levels \cite{fort2}.

On the other hand, the field of plasma physics provides an ideal platform to examine the fundamental theories at the most basic level due to accommodation of variety of collective electrostatic as well as electromagnetic wave-particle interactions. Due to this colorful interactions among the plasma species many interesting new physical phenomena can be probed in ionized environments such as the wide range of electromagnetic propagation modes \cite{se}, a variety of interesting plasma instabilities \cite{bun} and many astonishing nonlinear phenomena such as the multi-wave interactions \cite{sten}, raman scattering \cite{ses}, charge screening, pondermotive effect, higher harmonic generation, sheath formation, nonlinear wave resonance and bistability \cite{akb1}, chaos and turbulence, etc. to name a few. Quantum plasmas has evolved since the early development date by pioneering contributions of many physicists like Fermi \cite{fermi}, Madelung \cite{madelung}, Hoyle and Fowler \cite{hoyle}, Chandrasekhar \cite{chandra}, Bohm \cite{bohm}, Pines \cite{pines}, Levine \cite{levine}, Klimontovich and Silin \cite{klimontovich} and many others. Recently, there has been a renewed interest in the field due to the wide range of relevant applications in the rapidly growing interdisciplinary areas of optoelectronics and semiconductor devices \cite{hu,seeg}, photonic crystals \cite{mark}, nanofabrication technology, plasmonics \cite{man1}, and many others. Recent theoretical and numerical investigations on quantum plasmas \cite{man2,shuk1,manfredi,haas1,brod1,mark1,man3,mold2,sm} with the help of fast and accurate computing technology have also revealed many unpredicted collective quantum aspects of the physical phenomena which may completely replenish and revolutionize our static view of the old modern physics. In this paper we study the interaction plasmon excitations with a periodic crystal most relevant to solid state plasmas and provide an energy band theory based on a new concept of pseudo-resonance. We also revisit the historic problem of wave-particle duality in the context collective quantum interactions using the more generalized Schr\"{o}dinger-Poisson model which takes into account both single as well as the collective interactions in plasmas. The paper is organized as follows. We provide the hydrodynamic model for plasmon evolution in a fixed ion lattice in Sec. II. The equivalent Madelung Schr\"{o}dinger-Poisson system is introduced in Sec. III. We also introduce the pseudoforce and pseudo-resonance effect as the equivalent concept for energy band gap for solid states in Sec. IV. The wave-particle duality of plasmons is discussed in Sec. V. The energy band structure of plasmons is provided in Sec VI. The generalized solution with arbitrary periodic potential plasmonic crystal is given in Sec. VII and conclusions are drawn in Sec VIII.

\section{Hydrodynamic Model}

Let us consider a one dimensional (1D) free electron gas with arbitrary degree of degeneracy within an ideal periodic arrangement of singly ionized ion or other potential lattice. The latter may be considered as a simple physical model of 1D metal, semiconductor or plasmonic crystal with the electron gas chemical potential, $\mu_0$, and temperature, $T$. Then, the evolution of the nearly free plasmon inside the ambient periodic potential may be fully described by the hydrodynamic model which constitutes the complete set of continuity, fluid momentum balance and Poisson's equations as follows
\begin{subequations}\label{hd}
\begin{align}
&\frac{{\partial n}}{{\partial t}} + \frac{{\partial nu}}{{\partial x}} = 0,\\
&m\left[ {\frac{{\partial u}}{{\partial t}} + u\frac{{\partial u}}{{\partial x}}} \right] = e\frac{{\partial \phi }}{{\partial x}} - \frac{{\partial \mu }}{{\partial x}} + \frac{{{\hbar ^2}}}{{2m}}\frac{\partial }{{\partial x}}\left( {\frac{1}{{\sqrt n }}\frac{{{\partial ^2}\sqrt n }}{{\partial {x^2}}}} \right),\\
&\frac{{{\partial ^2}\phi }}{{\partial {x^2}}} = 4\pi e({n_e} - {n_i}),
\end{align}
\end{subequations}
in which $n$, $u$ and $m$ are the electron number density, fluid velocity and mass, respectively. Also, $\phi$, $\mu$ and $n_i$ are the electrostatic potential, chemical potential of electron gas and ion number density, respectively. The isothermal equation of state for electron fluid is given as
\begin{equation}\label{pol}
{n(\mu_0,T)} =  - {N}{\rm{L}}{{\rm{i}}_{3/2}}\left[ { - {\rm{exp}}\left( {-\beta {\mu_0}} \right)} \right],\hspace{3mm}{P(\mu_0,T)} =  - \frac{N}{\beta}{\rm{L}}{{\rm{i}}_{5/2}}\left[ { - {\rm{exp}}\left( {-\beta {\mu_0}} \right)} \right],
\end{equation}
in which $\beta=1/k_B T$, $k_B$ is the Boltzmann constant and $N$ is defined as
\begin{equation}\label{N}
{N} = \frac{2}{{\Lambda^3}} = 2\left( {\frac{{{m k_B T}}}{{{2\pi\hbar^2}}}} \right)^{3/2},
\end{equation}
with $\Lambda$ being the thermal de Broglie wavelength for electron gas. Also, the function ${\rm{Li}}_\nu(z)$ in (\ref{pol}) is the polylogarithm function of the order $\nu$ with the argument $z$ which is related to the Fermi integrals.

\section{The Schr\"{o}dinger-Poisson model}

The hydrodynamic model (\ref{hd}) may be compactly represented as the following effective Schr\"{o}dinger-Poisson model \cite{manfredi}
\begin{subequations}\label{sp}
\begin{align}
&i\hbar \frac{{\partial \cal N }}{{\partial t}} =  - \frac{{{\hbar ^2}}}{{2m}}\frac{{\partial ^2 {{\cal N}}}}{{\partial {x^2}}} - [e\phi-\mu(n,T)]{\cal N},\\
&\frac{{\partial {\phi ^2}}}{{\partial {x^2}}} = 4\pi e (n-n_i),
\end{align}
\end{subequations}
where ${\cal N} =\sqrt{n(x,t)}\exp[iS(x,t)]$ is the electron fluid wave function with ${\cal N}{\cal N^*}=n(x,t)$ being the number density and $u(x,t)=(1/m)\partial S(x,t)/\partial x$ fluid speed of the arbitrary degenerate electron gas in the hydrodynamic representation (\ref{hd}). The small amplitude perturbation of the free electron gas can be described using the appropriately linearized system assuming that the function ${\cal N}$ can be decomposed as ${\cal N}(x,t)=\psi(x)\exp(-i\epsilon t/\hbar)$ possessing energy eigenvalues, $\epsilon$
\begin{subequations}\label{pf}
\begin{align}
&\frac{{d^2{\Psi(x)}}}{{d{x^2}}} + \Phi(x) = - 2 E \Psi(x),\\
&\frac{{d^2{\Phi(x)}}}{{d{x^2}}} - {\Psi}(x) = 0,
\end{align}
\end{subequations}
in which the normalization scheme is used as follows. The generalized wave function is defined as $\Psi=\psi/\sqrt{n_0}$ with $n_0$ being the equilibrium number density of electron gas and the electrostatic potential function, $\Phi=e\phi$, with $E=(\epsilon-\mu_0)/2E_p$ where $E_p=\hbar\omega_p$ with $\omega_p=\sqrt{4\pi e^2 n_0/m}$ is the plasmon energy quanta and $x=x/\lambda_p$ where $\lambda_p=2\pi/k_p$ with $k_p=\sqrt{2m E_p}/\hbar$ is a very important characteristic plasma length, so called, the plasmon wavelength which characterizes the plasma interaction range. The later may also perfectly characterize the wave-particle duality of plasmon excitations in the quantum physics as is illuminated in the proceeding. Note also that $\mu_0$ is the equilibrium chemical potential of the electron gas. Therefore, by ignoring the change in the chemical potential of the electron gas, current theory does not apply to the large amplitude nonlinear or turbulent evolution of the free electron gas. The model (\ref{pf}) can now be generalized to include the periodic lattice potential as a pseudoforce potential of the form $U(x)$. Therefore we have
\begin{subequations}\label{gpf}
\begin{align}
&\frac{{d^2{\Psi(x)}}}{{d{x^2}}} + \Phi(x) = - 2 E \Psi(x),\\
&\frac{{d^2{\Phi(x)}}}{{d{x^2}}} - {\Psi}(x) = U\cos(G x),
\end{align}
\end{subequations}
in which the lattice potential is contributed by an approximate cosine potential of the form $U(x)=U\cos(Gx)$ where $U$ is a constant average amplitude of lattice potential and $G$ is the reciprocal lattice vector of the 1D crystal with a period $a$, hence, $G=2\pi/a$. It is now clear that the model (\ref{gpf}) constitutes a simple coupled linear driven pseudoforce system.

\section{Driven Pseudoforce System and Pseudo-Resonance}

To this end we present the general solution of the system (\ref{gpf}) as follows \cite{akb2}
\begin{subequations}\label{sol1}
\begin{align}
&\Phi (x) = \frac{U}{{2\alpha }}\left[ {\frac{{\left( {1 + \beta {k_2^2}} \right)\cos \left( {{k_1}x} \right)}}{{\left( {{G^2} - {k_1^2}} \right)\left( {{G^2} - {k_2^2}} \right)}} - \frac{{\left( {1 + \beta {k_1^2}} \right)\cos \left( {{k_2}x} \right)}}{{\left( {{G^2} - {k_1^2}} \right)\left( {{G^2} - {k_2^2}} \right)}} - \frac{{2\alpha \beta \cos \left( {Gx} \right)}}{{\left( {{G^2} - {k_1^2}} \right)\left( {{G^2} - {k_2^2}} \right)}}} \right]\\
&+ \frac{1}{{2\alpha }}\left[ {\left( {{k_2^2}{\Phi _0} + {\Psi _0}} \right)\cos \left( {{k_1}x} \right) - \left( {{k_1^2}{\Phi _0} + {\Psi _0}} \right)\cos \left( {{k_2}x} \right)} \right],\\
&\Psi (x) = \frac{U}{{2\alpha }}\left[ {\frac{{\cos \left( {{k_2}x} \right)}}{{\left( {{G^2} - {k_2^2}} \right)}} - \frac{{\cos \left( {{k_1}x} \right)}}{{\left( {{G^2} - {k_1^2}} \right)}} - \frac{{2\alpha \cos \left( {Gx} \right)}}{{\left( {{G^2} - {k_1^2}} \right)\left( {{G^2} - {k_2^2}} \right)}}} \right]\\
&+ \frac{1}{{2\alpha }}\left[ {\left( {{\Phi _0} + {k_2^2}{\Psi _0}} \right)\cos \left( {{k_2}x} \right) - \left( {{\Phi _0} + {k_1^2}{\Psi _0}} \right)\cos \left( {{k_1}x} \right)} \right].
\end{align}
\end{subequations}
where $\Phi_0$ and $\Psi_0$ are arbitrary boundary values and
\begin{equation}\label{const1}
\alpha  = \sqrt {{E^2} - 1},\hspace{3mm}\beta  = {G^2} - 2E,\hspace{3mm}{k_1} = \sqrt {E - \alpha },\hspace{3mm}{k_2} = \sqrt {E + \alpha }.
\end{equation}
It is easily confirmed that $k_1$ and $k_2$ are reciprocal of each other, $k_1 k_2=1$. For periodic solutions which satisfy the lattice periodicity we require that
\begin{subequations}\label{bc1}
\begin{align}
&\Phi (0) = \Phi (na),\hspace{3mm}{\left. {\frac{{d\Phi (x)}}{{dx}}} \right|_{x = 0}} = {\left. {\frac{{d\Phi (x)}}{{dx}}} \right|_{x = na}},\\
&\Psi (0) = \Psi (na),\hspace{3mm}{\left. {\frac{{d\Psi (x)}}{{dx}}} \right|_{x = 0}} = {\left. {\frac{{d\Psi (x)}}{{dx}}} \right|_{x = na}},
\end{align}
\end{subequations}
where $n$ is an integer number and $a$ is the lattice period. The four bounady conditions (\ref{bc1}) consistently give the following relations for initial values
\begin{equation}\label{zeros}
{\Phi _0} =  - \frac{U{\left( {{n^2}{G^2} - {k_1^2} - {k_2^2}} \right)}}{{\left( {{n^2}{G^2} - {k_1^2}} \right)\left( {{n^2}{G^2} - {k_2^2}} \right)}},\hspace{3mm}{\Psi _0} =  - \frac{U}{{\left( {{n^2}{G^2} - {k_1^2}} \right)\left( {{n^2}{G^2} - {k_2^2}} \right)}},
\end{equation}
where $G=2\pi/a$ is the smallest reciprocal lattice vector. Plugging (\ref{zeros}) into (\ref{sol1}) gives the particular solutions as
\begin{equation}\label{psol1}
\Phi_n(x) = {\Phi _0}\cos (nGx),\hspace{3mm}\Psi_n(x) = {\Psi _0}\cos (nGx).
\end{equation}
Note that the presence of $n$ in the solutions(\ref{psol1}) only indicates that the solutions apply to the sublattice with a general reciprocal vector $nG$ in order to contribute to multiple Brillouin zones. Hence, in what follows we focus on the case $n=1$ in the reduced zone scheme.

Figure 1 depicts the solution profiles for $\Psi$ and $\Phi$ for tentative values of normalized plasmon energy $E=(\epsilon-\mu_0)/2\epsilon_p$, lattice potential strength parameter $U$ and lattice space parameter $a$. Figure 1(a) shows the wave function $\Psi(x)$ profile (thick curve) and the cosine lattice potential (thin curve) with filled black circles denoting the lattice ion positions. It is noted that the two periodic profiles are $180$ degrees out of phase, so that, the electron gas density is localized at the position of lattice ions and ${\cal  N}$ denotes a standing wave. Figure 1(b) shows the electrostatic potential profile due to collective plasmon excitation which is in phase with the lattice potential. Figure 1(c) shows $\Psi(x)$ for a different value of $E$ with the same values for $U$ and $a$. It is observed that for this value of $E$ the profile for $\Psi$ is in phase with the lattice potential. It is also noted from Fig. 1(d) that for the same energy value as Fig. 1(c) the potential profile $\Phi$ is $180$ degrees out of phase with the lattice potential.

A closer look at the particular solutions (\ref{psol1}) reveals that they are pseudo-resonant solutions with double pseudo-resonance values of $k_1,k_2=G$. Since, $\Psi\Psi^*$ represents the electron gas number density,in the energy profiles of Fig. 1(a) and 1(c) electrons must be localized on ions or the midway between the ions in the standing wave configuration. In the reciprocal space language the resonant condition occur when either of the wavevectors $k_1$ or $k_2$ end at the Brillouin zone $k_B=\pm n\pi/a$ \cite{kit,ash}. Figure 2(a) shows  the real space regions for positive/negative (colored/white) phase shifts between $\Psi(x)$ and $U(x)$. The border between these regions indicates the resonant values for $E$ and $a$. Figure 2(b) shows the positive/negative (colored/white) phase shifts between $\Psi(x)$ and $U(x)$ in reciprocal space. It is remarked that Phase shift in real space is always opposite to that in $k$-space. t is also revealed that in both Fig. 2(a) and 2(b) there is a minimum value of $E=2E_p$ below which there is no pseudo-resonant condition. In fact the energy gap of $2 E_p$ corresponds to plasmon propagation itself, that is, no plasmon with energy below $\epsilon_g=\mu_0+2E_p$ can exist in the free electron gas with a chemical potential of $\mu_0$ \cite{kit}. Figure 2(c) shows the variation in amplitude ($\Psi_0$) of $\Psi(x)$ for different values of normalize energy $E$ in terms of the normalized lattice spacing. It is revealed that the amplitude $\Psi_0$ diverges for $k=\pi/a$ and $k=a/\pi$. The condition $k=\pi/a$ indicates the normal first energy gap in a 1D solid \cite{kit,ash}. The other pseudo-resonance condition $k=a/\pi$ is completely new condition for the energy gap value for the 1D plasmonic crystal such as solids \cite{kit,ash}. However, each type of energy gap occur for lattice spacings $a<2\pi\lambda_p$ or $a>2\pi \lambda_p$. The typical values of lattice constant, $a$, for most metallic elements is below the critical value of $a_c=2\pi\lambda_p$ \cite{kit,ash}, hence, the energy gap for them is ordinary one. Meanwhile, for artificial plasmonic crystals with $a>2\pi\lambda_p$ the second types of energy gaps with $k=a/\pi$ may form. Form Fig. 2(c) it is remarked that the energy gap for larger $a>a_c$ are relatively bigger that those for $a<a_c$. Figure 2(d) shows the variation in amplitude ($\Psi_0$) of $\Psi(x)$ for different values of normalize energy $E$ in terms of the normalized reciprocal lattice vector, $G$. It is observed from this plot that the energy gap openings for smaller reciprocal lattice vectors \cite{kit,ash} are larger compared to that of larger reciprocal lattice vectors and viceversa.

Figure 3 shows the qualitative variation in the first band gap structure for plasmonic crystals with different lattice constants. It is observed from Fig. 3 that both wave function and electrostatic potential amplitudes diverge at identical values of the normalized energy $E$. Figure 3(a) and 3(b) indicate that for over critical lattice constants ($a>a_c=2\pi\lambda_p$) the band gap in $E$ shifts to higher values of the energy as the lattice constant increases. On the other hand, for under critical values of lattice constant ($a<a_c=2\pi\lambda_p$) the energy gap value shifts to lower energy as the lattice constant increases as oposed to the over critical lattice spacing. The normalized dispersion relation for plasmon excitations can be found from (\ref{const1}) to be $E=(1+k^4)/2k^2$. This is obviously different from the parabolic dispersion of single electron in a free electron gas. It is remarked that the presence of collective effects radically modifies the energy dispersion relation due to plasmon excitations.

\section{Plasmon Wave-Particle Duality}

In figure 4(a) we have depicted the scaled ($\hbar=1$) phase $\omega/k$ (dash-dotted) and group $d\omega/dk$ (dashed) speeds as well as the effective mass $1/m^*=(1/\hbar^2)d^2E/dk^2$ (solid) for the free electron parabolic dispersion relation. It is remarked that both phase and group speeds vary linearly with the wavenumber and the relation $V_{p}=V_g/2$ always holds. Also, the effective  mass is always equal to the electron mass. Moreover, figure 4(b) shows these parameters values for plasmon dispersion relation given above. While, for the liming regime $k\to\infty$ the values for plasmon approached that of the free electron, the parameter values are radically different for $k<k_p$ or equivalently for the length scales $l>\lambda_p$. The most interesting feature of plasmon is that the fractional mass $m^*/m$ approaches the value of zero for the limit of $k\to 0$ or $\lambda\to\infty$. It is remarked that for $k\gg k_p$ ($l\ll \lambda_p$) the particle aspects of plasmon appears, whereas, for $k\ll k_p$ ($l\gg \lambda_p$) the wave features with zero mass take place (the Bohr's complementarity principle \cite{bohr}). The later is clearly a remarkable wave-particle duality manifestation of the plasmon excitations which has intrigued the scientists since the very birth of the quantum mechanics with almost no clear physical explanation of it. It seems however that the plasmon itself is the de Broglie-Bohm pilot wave \cite{de}. Note that the complementarity principle may break down for length scales comparable to that of the plasmon excitation which lies at the nanoscale region for most metallic compounds.

\section{Plasmon Energy Band Structure}

Figure 5(a) depicts different branches of the plasmon excitation energy dispersion which follows the relations (\ref{const1}). The dispersion has distinct minimum values at the value of $k_1=k_2=\pm k_p$. Now using the identity $k_1=1/k_2$ the dispersion of plasmon excitation reduces to $E=(1+k^4)/2k^2$ which is compared with the parabolic energy dispersion of free single electron \cite{kit,ash} in Fig. 5(b). It is remarked that for $k\gg k_p$ or the length scales much smaller than the plasmon wavelength, where the particle aspects emerge, the two dispersion relations match approach one another. The later applies to the atomic angstrom scales where the single electron interacts with the atoms. On the other hand, the plasmon dispersion is seen to completely deviate from the single electron one for $k\ll k_p$ which is the super nano-length scales, where completely wave aspects are detectable. The later applies to electronic wave function in the much larger classical metallic scales. Moreover, figure 5(c) depicts the free electron dispersion in the extended zone scheme \cite{kit,ash} with lattice parameter value of $a=\pi$. This scheme approximately describes the energy (band) gap positions for simple metallic elements as the places in which dispersion curves from different Brillouin zones intersect at the zone boundary, $k=\pm n\pi/a$. Figure 5(d) shows both the plasmon and the free electron energy dispersion curves in a single plot for the same value of lattice parameter, $a$. It is seen that there is large band gap for $E<2E_p$ for which there is no plasmon excitations. This is in sharp contrast with the conventional free electron extended zone profile as shown with the dashed curves in this plot. For higher band gaps, however, there is a relative match for the two theories with the gap positions in the plasmon dispersion scheme being slightly lower compared to those of the free electron. The match off course become better as the energy of the bands increase. It is also remarked that for this under critical lattice spacing $a< 2\pi\lambda_p$ the wave branches ($1/2k^2$) in the plasmon dispersion do not lead to extra energy gaps.

Figure 6 shows the energy band gap structure for single free electrons and plasmon excitations for over- and under-critical lattice constant in the reduced scheme, $-\pi/a<k<\pi/a$. Comparison of Figs. 6(a) and 6(b) shows that in the free electron model the energy bands become denser as the lattice parameter $a$ increases. Figures 6(c) and 6(d), on the other hand, depict the plasmon dispersion energy bands for the same parameter values in the free electron band structure \cite{kit,ash} of Figs. 6(a) and 6(b). It is clearly remarked that, while the band structure at under-critical lattice parameter value for free electron and current plasmon model nearly match, for the over-critical lattice constant values, $a>2\pi\lambda$, extra bands appear with new gap positions. It is also remarked that some degenerate gap develop $4<E/2E_p<5$ due to overlap of the $1/2k^2$ branch of the plasmon dispersion. The later is completely a new feature of energy band gap in solids with over-critical atomic spacing applicable to artificial plasmonic crystals.

\section{Plasmonic Lattice with Arbitrary Periodic Potential}

Let us first consider the 1D crystal model with period $a$ and two atoms in a unit cell one located at the origin and the other at $x=b$. The lattice potential is then approximated as $U(x)=A\cos(Gx)+B\cos[G(x-b)]$ in which $G=2\pi/a$ is the reciprocal space vector. Now using the same procedure and boundary condition as before and ignoring the long expressions for the general solutions $\Psi(x)$ and $\Phi(x)$, one obtains for $n=1$
\begin{equation}\label{zeros2}
{\Phi _0} =  - \frac{{(A+B)\left( {{G^2} - {k_1^2} - {k_2^2}} \right)}}{{\left( {{G^2} - {k_1^2}} \right)\left( {{G^2} - {k_2^2}} \right)}},\hspace{3mm}{\Psi _0} =  - \frac{(A+B)}{{\left({{G^2} - {k_1^2}} \right)\left( {{G^2} - {k_2^2}} \right)}},
\end{equation}
where plugging (\ref{zeros2}) into general solutions gives the particular solutions as
\begin{equation}\label{psol2}
\Phi (x) = \frac{{{\Phi _0}\{ A\cos (Gx) + B\cos [G(x - b)]\} }}{{A + B}},\Psi (x) = \frac{{{\Psi _0}\{ A\cos (Gx) + B\cos [G(x - b)]\} }}{{A + B}}.
\end{equation}
Figure 7 shows the wave function and electrostatic potential profiles for particular solutions (\ref{psol2}) (thick curves) along with the lattice potential of each atom separately (thin curves) for given tentative solution parameter values, indicated above each plot. The position of different atoms are shown with filled/empty circles. The Fig. 7 shows by comparison the same features as Fig. 1 in which there is a $180$ phase shift on increasing of the normalized plasmon energy between the total lattice potential and the wave function profile, as confirmed in the solution (\ref{psol2}). It is remarked that the band structure of 1D plasmonic crystal is completely independent of the parameter $b$ and the number of atoms in the unit cell. Also, note the remarkable property of the solutions (\ref{psol2}) in which they are linear composition of single atom sublattice solutions.

To this end let us now consider the lattice with two reciprocal space vectors $G_1$ and $G_2$. It can be shown that the solution to this case also decomposes to sublattice solution if we have $G_1$ and $G_2$ have the same period as is the case for all crystals with periodic structure. Therefore, we would like to consider the most general periodic potential $f(x)=f(x+na)$ for the lattice ion with period $a$. Using the Fourier expansion of the lattice potential in terms of the first reciprocal lattice vector we obtain
\begin{equation}\label{gpot}
f(x) = \sum\nolimits_n {{C_n}\exp \left( {inGx} \right)},\hspace{3mm}{C_n} = \frac{1}{a}\int\limits_0^a {f(x)\exp \left( { - inGx} \right)dx},
\end{equation}
in which $n$ is the integer value summation index which varies in the range $-\infty<n<+\infty$ and $C_n$ are the Fourier components of the potential. Now defining the general notation $G_n=nG=2n\pi/a$ and $C_G=C_n$ we may write the general particular solutions for the plasmon wave function and electrostatic potentials in a generalized plasmonic lattice potential as
\begin{equation}\label{psolg}
\Phi (x) =  - \sum\limits_{ - \infty }^{ + \infty } {\frac{{{C_G}\left( {G_n^2 - k^2 - k^{-2}} \right)\exp \left(i{G_n}x\right)}}{{\left( {G_n^2 - k^2} \right)\left( {G_n^2 - k^{-2}} \right)}}},\hspace{3mm}\Psi (x) =  - \sum\limits_{ - \infty }^{ + \infty } {\frac{{{C_G}\exp \left(i{G_n}x\right)}}{{\left( {G_n^2 - k^2} \right)\left( {G_n^2 - k^{-2}} \right)}}}.
\end{equation}
Note that the infinite number of pseudo-resonance conditions in (\ref{psolg}) attribute to infinite number of the Brillouin zones of plasmonic crystal in the extended zone representation.

\section{Conclusion}

In this paper we studied the energy band structure of a plasmonic crystal. Using the Schr\"{o}dinger-Poisson model we deduced the appropriate generalized coupled pseudoforce model with the periodic lattice potential as the driving force. The pseudo-resonance condition leads to the energy gaps at the plasmon dispersion which is found to be quite different from that of free electron one. We found an important characteristic length for the critical lattice parameter above and below which the energy band structure of lattice becomes fundamentally different. The energy dispersion of the plasmon clearly explains wave-particle duality of quantum particles which have intrigued the scientists since the start of the quantum theory. Current development helps to better understand collective effects in quantum phenomenon to the root level. Current theory can also have a wide application in the rapidly growing nanoelectronics, plasmonics and solid state.


\begin{thebibliography}{}
\bibitem {planck} M. Planck, Ann. Physik \textbf{1}, 73(1900)
\bibitem {es} E. Schr\"{o}dinger, Phys. Rev. \textbf{28}, 1049(1926)
\bibitem {hb} W. Heisenberg, Zeitschrift für Physik (in German), \textbf{43} 172(1927).
\bibitem {en} A. Einstein (1916), Relativity: The Special and General Theory (Translation 1920), New York: H. Holt and Company
\bibitem {born1} M. Born, Zeitschrift für Physik, \textbf{37} (12) 863(1926).
\bibitem {born2} Max Born and Emil Wolf, Principles of Optics, Cambridge University Press (1999).
\bibitem {bohr} N. Bohr, Dialectica \textbf{2} 312(1948).
\bibitem {dg} C. Davisson, and L. H. Germer, Phys. Rev. Band \textbf{30}, 705(1927).
\bibitem {de} L. de Broglie, Ann. de la Fondation \textbf{12}, 1(1987)
\bibitem {born3} M. Born, Nature \textbf{119} 354(1927).
\bibitem {bohm1} D. Bohm, Phys. Rev. \textbf{85} 166(1952); doi:10.1103/PhysRev.85.166
\bibitem {co1} Y. Couder and E. Fort, Phys. Rev. Lett. \textbf{97}, 154101(2006).
\bibitem {co2} Y. Couder, A. Boudaoud, S. Proti‘ere, and E. Fort, Europhysics News \textbf{41}, 14(2010).
\bibitem {bush} J. W. M. Bush, Proceedings of the National Academy of Sciences \textbf{107}, 17455(2010).
\bibitem {co3} Y. Couder, S. Proti‘ere, E. Fort, and A. Boudaoud, Nature \textbf{437}, 208208(2005).
\bibitem {co4} Y. Couder and E. Fort, Journal of Physics: Conference Series \textbf{361}, 012001(2012)
\bibitem {eddi} A. Eddi, E. Fort, F. Moisy, and Y. Couder, Phys. Rev. Lett. \textbf{102}, 240401(2009).
\bibitem {fort1} E. Fort and Y. Couder, EPL (Europhysics Letters) \textbf{102}, 16005(2013).
\bibitem {fort2} E. Fort, A. Eddi, A. Boudaoud, J. Moukhtar, and Y. Couder, Proceedings of the National Academy of Sciences \textbf{107}, 17515(2010).
\bibitem {se} P. K. Shukla and B. Eliasson Phys. Rev. Lett. \textbf{99}, 096401(2007).
\bibitem {bun} O. Buneman, Physical Review Letters, \textbf{10}, 285(1963).
\bibitem {sten} L Stenflo Phys. Scr. \textbf{1994T50} 15(1994).
\bibitem {ses} P. K. Shukla, B. Eliasson, and L. Stenflo Phys. Rev. E \textbf{86}, 016403(2012).
\bibitem {akb1} M. Akbari-Moghanjoughi, Phys. Plasmas \textbf{24}, 082108(2017); doi.org/10.1063/1.4990458
\bibitem {fermi} E. Fermi and E. Teller, Phys. Rev. {\bf 72}, 399 (1947).
\bibitem {madelung} E. Madelung, Z. Phys., 40 322(1926).
\bibitem {hoyle} F. Hoyle and W. A. Fowler, Astrophys. J. \textbf{132}, 565(1960).
\bibitem {chandra} S. Chandrasekhar, "\emph{An Introduction to the Study of Stellar Structure}", The University of Chicago Press, Chicago (1939).
\bibitem {bohm} D. Bohm and D. Pines, Phys. Rev. \textbf{92} 609(1953).
\bibitem {pines} D. Pines, Phys. Rev. \textbf{92} 609(1953).
\bibitem {levine} P. Levine and O. V. Roos, Phys. Rev, \textbf{125} 207(1962).
\bibitem {klimontovich} Y. Klimontovich and V. P. Silin, in Plasma Physics, edited by J. E. Drummond (McGraw-Hill, New York, 1961).
\bibitem {hu} C. Hu, Modern Semiconductor Devices for Integrated Circuits (Prentice Hall, Upper Saddle River, New Jersey, 2010) 1st ed.
\bibitem {seeg} K. Seeger, Semiconductor Physics (Springer, Berlin, 2004) 9th ed.
\bibitem {mark} P. A. Markovich, C.A. Ringhofer, and C. Schmeister, Semiconductor Equations (Springer, Berlin, 1990).
\bibitem {man1} G. Mangredi, Phys. Plasmas \textbf{25}, 031701(2018); https://doi.org/10.1063/1.5026653
\bibitem {man2} G. Manfredi, “How to model quantum plasmas,” Fields Inst. Commun. {\bf 46}, 263–287 (2005); in Proceedings of the Workshop on Kinetic Theory (The
Fields Institute, Toronto, Canada 2004): http://arxiv.org/abs/quant--ph/0505004.
\bibitem {shuk1} P. K. Shukla, B. Eliasson, ”Nonlinear aspects of quantum plasma physics” Phys. Usp. \textbf{51}
53(2010).
\bibitem {manfredi} G. Manfredi and F. Haas, Phys. Rev. B {\bf 64}, 075316 (2001);
\bibitem {haas1} F. Haas, {\sl Quantum Plasmas: An Hydrodynamic Approach} (Springer, New York, 2011).
\bibitem {brod1} G. Brodin and M. Marklund, New J. Phys. \textbf{9}, 277(2007).
\bibitem {mark1} M. Marklund and G. Brodin, Phys. Rev. Lett. \textbf{98}, 025001(2007).
\bibitem {man3} N. Crouseilles, P. A. Hervieux, and G. Manfredi, Phys. Rev. B {\bf 78}, 155412 (2008).
\bibitem {mold2} Z. Moldabekov, Tim Schoof, Patrick Ludwig, Michael Bonitz, and Tlekkabul Ramazanov, Phys. Plasmas, \textbf{22}, 102104(2015); doi.org/10.1063/1.4932051
\bibitem {sm} L. Stanton and M. S. Murillo, Phys. Rev. E \textbf{91}, 033104(2015).
\bibitem {akb2} M. Akbari-Moghanjoughi, "Characteristics of Plasmon Transmittivity Over Potential Barriers" Submitted to Physics of Plasmas
\bibitem {kit} C. Kittel, Introduction to Solid State Physics, (John Wiely and Sons, New York, 1996), 7th ed.
\bibitem {ash} N. W. Ashcroft and N. D. Mermin, Solid state physics (Saunders College Publishing, Orlando, 1976).
\end{thebibliography}
\end{document}